\documentclass[aps,prl,twocolumn,superscriptaddress,floatfix,showpacs,nofootinbib]{revtex4-1}
\usepackage{epsfig}
\usepackage{dcolumn}
\usepackage{bm}
\usepackage{amssymb}
\usepackage{amsmath}
\usepackage{color}
\usepackage{subfigure}
\usepackage{lipsum}
\usepackage[bottom]{footmisc}

\newcommand{\be}{\begin{equation}}
\newcommand{\ee}{\end{equation}}
\newcommand{\bea}{\begin{eqnarray}}
\newcommand{\eea}{\end{eqnarray}}

\usepackage[normalem]{ulem}  

\renewcommand\sout{\bgroup \color{red} \ULdepth=-.5ex \ULset}

\begin{document}

\title{Probing the partonic degrees of freedom in high multiplicity p-Pb  collisions at $\sqrt{s_{NN}}=$ 5.02 TeV }

\author{Wenbin Zhao}
\affiliation{Department of Physics and State Key Laboratory of Nuclear Physics and Technology, Peking University, Beijing 100871, China}
\affiliation{Collaborative Innovation Center of Quantum Matter, Beijing 100871, China}

\author{Che Ming Ko}
\affiliation{Cyclotron Institute and Department of Physics and Astronomy, Texas A$\&$M University, College Station, TX 77843, USA}

\author{Yu-Xin Liu}
\affiliation{Department of Physics and State Key Laboratory of Nuclear Physics and Technology, Peking University, Beijing 100871, China}
\affiliation{Collaborative Innovation Center of Quantum Matter, Beijing 100871, China}
\affiliation{Center for High Energy Physics, Peking University, Beijing 100871, China}

\author{Guang-You Qin}
\affiliation{Institute of Particle Physics and Key Laboratory of Quark and Lepton Physics (MOE), Central China Normal University, Wuhan, Hubei 430079, China}
\affiliation{Nuclear Science Division, Lawrence Berkeley National Laboratory, Berkeley, CA, 94270, USA}

\author{Huichao Song}
\affiliation{Department of Physics and State Key Laboratory of Nuclear Physics and Technology, Peking University, Beijing 100871, China}
\affiliation{Collaborative Innovation Center of Quantum Matter, Beijing 100871, China}
\affiliation{Center for High Energy Physics, Peking University, Beijing 100871, China}

\date{\today}

\begin{abstract}
We investigate the role of partonic degrees of freedom in high multiplicity p-Pb collisions at $\sqrt{s_{NN}}=$ 5.02 TeV carried out at the Large Hadron Collider (LHC) by studying the production and collective flow of identified hadrons at intermediate $p_T$  via the coalescence of soft  partons  from the viscous hydrodynamics ({\tt VISH2+1}) and hard partons from the energy loss model ({\tt LBT}). We find  that combining  these intermediate $p_T$ hadrons  with the low $p_T$ hadrons from the hydrodynamically expanding fluid and high $p_T$ hadrons from the fragmentation of quenched jets, the resulting {\tt Hydro-Coal-Frag} model provides a nice description of measured $p_T$-spectra and differential elliptic flow $v_2(p_T)$ of pions, kaons and protons over the $p_T$ range from 0 to 6 GeV. We further demonstrate the necessity of including the quark coalescence contribution to reproduce the experimentally observed approximate number of constituent quark scaling of hadron $v_2$ at intermediate $p_T$. Our results thus indicate the importance of partonic degrees of freedom and  also hint at the possible formation of quark-gluon plasma in high multiplicity p-Pb collisions at the LHC.
\end{abstract}

\maketitle

\noindent \textsl{1. Introduction.} The search for the quark gluon plasma (QGP) and studying its properties are the main goals of relativistic heavy ion collisions during the past two decades. Since early experiments on Au-Au collisions at $\sqrt{s_{NN}}$= 200 GeV at the Relativistic Heavy Ion Collider (RHIC), many evidences have been accumulated for the formation of the QGP. Among them, the strong collective flow, the number of constituent quark (NCQ) scaling of elliptic flow ($v_2$) and the quenching of energetic jets are the three most prominent ones~\cite{Gyulassy:2004zy,Adcox:2004mh,Jacobs:2007dw,long1,Muller:2006ee}.

The p-Pb collisions at $\sqrt{s_{NN}}$= 5.02 TeV were originally aimed to study the cold nuclear matter effects and to provide the reference data for Pb-Pb collisions at the LHC. However, various striking features of collectivity have been observed in the high-multiplicity events of p-Pb collisions, including the long-range ``double ridge" structures in the 2-particle correlations~\cite{CMS:2012qk,Abelev:2012ola,Aad:2013fja,Khachatryan:2015waa}, the changing signs of the 4-particle cumulants~\cite{Aad:2013fja,Abelev:2014mda,Khachatryan:2015waa} and the mass ordering of the $v_2$ of identified hadrons~\cite{ABELEV:2013wsa,Khachatryan:2014jra}, etc.  This  has  also triggered  the study of collectivity for other small systems, such as p-Au, d-Au and $^3$He-Au  at RHIC~\cite{Aidala:2017ajz,PHENIX:2018lia,Mace:2018vwq,Schenke:2019pmk}  and p-p  at the LHC~\cite{Aad:2015gqa,Khachatryan:2016txc,ATLAS:2017tqk,Zhao:2017rgg,Acharya:2019vdf}, as well as the system size scan of various collision systems~\cite{Sievert:2019zjr,Citron:2018lsq,Katz:2019fkc,Katz:2019jcw}. The origin of the observed collective behavior  in small systems are still under
debate~\cite{Li:2012hc,Dusling:2015gta,Song:2017wtw,Nagle:2018nvi}. For the soft hadron data measured in  high-multiplicity p-Pb collisions, studies using hydrodynamics or transport models based on final-state effects can quantitatively or semi-quantitatively describe many of these flow-like signals~\cite{Werner:2013ipa,Bozek:2013ska,Shen:2016zpp,Schenke:2014zha,Mantysaari:2017cni,Qin:2013bha,Zhou:2015iba,Weller:2017tsr,Bozek:2015swa,Schenke:2019pmk,Bzdak:2014dia,Kurkela:2018ygx,Sun:2019gxg}. On the other hand, the color glass condensate (CGC)~\cite{Kovner:2012jm,Dusling:two,Dusling:2017dqg,Mace:2018yvl,Kovchegov:2012nd,Lappi:2015vta,Zhang:2019dth} and IP-Glasma models~\cite{Schenke:2015aqa,Schenke:2016lrs}, which  focus on the initial-state effects, have also been used to explain some features of the observed collectivity, but without an unambiguous or unquestionable conclusion. As to the hard probes, the energy lost by energetic partons can no longer lead to obvious signatures at high $p_T$ to discern if the QGP is formed or not due to its limited size and lifetime. The relatively small nuclear modification effects for large $p_T$ light and heavy flavor hadrons and jets measured in p-Pb collisions  at $\sqrt{s_{NN}}$= 5.02 TeV are consistent with the expectations of cold nuclear matter effects and very little hot medium effects~\cite{ATLAS:2014cza,ALICE:2012mj,Khachatryan:2015xaa,Acharya:2018qsh,Acharya:2017okq,Albacete:2013ei,Eskola:2016oht,
Abelev:2014hha,Khachatryan:2015uja,Dong:2019byy,Khachatryan:2015sva,Adam:2016ich,Du:2018wsj}.

Recently, the ATLAS, CMS and ALICE Collaborations have measured the   $v_2(p_T)$ of  charged and identified hadrons with high precision in the high multiplicity events of p-Pb collisions at $\sqrt{s_{NN}}$= 5.02 TeV~\cite{Sirunyan:2018toe,Aaboud:2016yar,Pacik:2018gix}. The resulting data shows a similar approximate NCQ scaling of $v_2$ at intermediate $p_T$ as observed in heavy ion collisions where the QGP is formed. In the present  work, we study the role of partonic degrees of freedom in these collisions and show that the observed approximate NCQ scaling of $v_2$ at intermediate $p_T$ can be explained if one includes the contribution to hadron production from the coalescence of thermal partons from the hydrodynamic fluid and the hard partons from jets.  Our study thus  demonstrates for the first time the importance of partonic degrees of freedom and hints  at  the possible formation of QGP in high multiplicity p-Pb collisions at the LHC.

\noindent \textsl{ 2. Methodology.}   Our study is based on the quark coalescence model that includes thermal-thermal, thermal-hard and hard-hard partons recombinations with thermal partons generated from the {\tt VISH2+1} hydrodynamics~\cite{Song:2007ux} and hard partons obtained from the {{\tt LBT}} energy loss model~\cite{Wang:2013cia,He:2015pra,Cao:2016gvr,Cao:2017hhk,Chen:2017zte,Luo:2018pto,Xing:2019xae}. In this paper,  we also develop a hybrid model, called {\tt Hydro-Coal-Frag}, that combines hydrodynamics at low $p_T$, coalescence at intermediate $p_T$ and fragmentation at high $p_T$, to study the  production and flow of identified hadrons for  $p_T <  6 \ \mathrm{GeV}$ in the high multiplicity events of p-Pb collisions.

In the coalescence model, the momentum distributions of mesons and baryons are calculated from the quark and antiquark phase-space distribution functions $f_{q,\bar q}({\bf x},{\bf p})$~\cite{Han:2016uhh}:
\begin{eqnarray}
\label{Eq1}
\frac{dN_M}{d^3 {\mathbf P}_M} &=& g_M\int d^3{\bf x}_1 d^3{\bf p}_1 d^3{\bf x}_2
d^3{\bf p}_2 f_{q}({\bf x}_1, {\bf p}_1) f_{\bar{q}}({\bf x}_2, {\bf p}_2)\nonumber\\
&&\times W_{M}({\bf y}, {\bf k})\delta^{(3)}({\bf P}_{M}-{\bf p}_1 -{\bf p}_2) \, ,\nonumber\\
 \label{Eq2}
\frac{dN_B}{d^3 {\bf P}_B} &=& g_B\int d^3{\bf x}_1 d^3{\bf p}_1 d^3{\bf x}_2 d^3{\bf p}_2 d^3{\bf x}_3 d^3{\bf p}_3 f_{q_1}({\bf x}_1, {\bf p}_1)\nonumber\\
&&\times f_{q_2}({\bf x}_2, {\bf p}_2)f_{q_3}({\bf x}_3, {\bf p}_3)W_{B}({\bf y}_{1},{\bf k}_{1};{\bf y}_{2},{\bf
k}_{2})\nonumber\\
&&\times \delta^{(3)}({\bf P}_{B}-{\bf p}_1 -{\bf p}_2 -{\bf p}_3).
\end{eqnarray}
Here, $g_{M,B}$ is the statistical factor for forming a meson or baryon of certain spin
, e.g., $g_{\pi}=g_K=\frac{1}{36}$ and $g_N=\frac{1}{108}$~\cite{Greco:2003xt}, and  $W_{M,B}$ is the smeared Wigner function of mesons or baryons. Following~\cite{Han:2016uhh}, the Wigner function of a meson in the $n$-th excited state is given by $W_{M,n}(\bf{y},\bf{k})=\frac{\emph{v}^n}{n!}e^{-\emph{v}}$ with $\emph{v}=\frac{1}{2}\left(\frac{{\bf y}^2}{\sigma_M^2}+{\bf k}^2 \sigma_M^2 \right)$, where ${\bf y}$ and ${\bf k}$ are the relative coordinates and momenta between the two constituent quark and antiquark in the meson.  The Wigner function of a baryon in the $n_1$-th and $n_2$-th excited states is given by $W_{B,n_1,n_2} (\mathbf{y}_1 , \mathbf{k}_1 ; \mathbf{y}_2, \mathbf{k}_2)=\frac{\emph{v}_1^{n_1}}{n_1!}e^{-\emph{v}_1}\cdot\frac{\emph{v}_2^{n_2}}{n_2!}e^{-\emph{v}_2}$, where $\emph{v}_i=\frac{1}{2}\left(\frac{{\bf y}_{i} ^2}{\sigma_{Bi}^2}+{\bf k}_{i} ^2 \sigma_{B_i^2}\right) $  with ${\bf y}_i$ and ${\bf k}_i$ being the relative coordinates and momenta among the three constituent quarks in the baryon. In this work, we include excited meson states up to $n=10$ and excited baryon states up to $n_1+n_2=10$~\cite{Han:2016uhh}. The width parameters $\sigma_M$, $\sigma_{B1}$ and $\sigma_{B2}$ are determined by the radii of formed hadrons~\cite{Beringer:1900zz}.

For the phase-space distributions $f_{q,\bar q}({\bf x},{\bf p})$, they are taken from the {\tt VISH2+1} hydrodynamics~\cite{Song:2007ux} for thermal partons and from the {{\tt LBT}} energy loss
model~\cite{Wang:2013cia,He:2015pra,Cao:2016gvr,Cao:2017hhk,Chen:2017zte,Luo:2018pto,Xing:2019xae} for hard partons.  {{\tt VISH2+1}} is a 2+1-dimensional viscous hydrodynamic code that simulates the expansion of the QGP fireball. Following~\cite{Zhao:2017yhj,Zhao:2017rgg,Zhao:2018lyf,Lming}, we use the s95-PCE~\cite{Huovinen:2009yb,Shen:2010uy} equation of state and the {{\tt TRENTo}} initial conditions~\cite{Bernhard:2016tnd,Moreland:2018gsh} with a starting time $\tau_0=0.8$ fm/c.  For the specific shear and bulk viscosity, their values are taken from Refs.~\cite{Ryu:2015vwa,Denicol:2009am,Bernhard:2016tnd}.   With such  parameter set,  {{\tt VISH2+1}} can  reproduce the measured $p_T$-spectra and $v_2(p_T)$ of identified hadrons for $p_T<2$ GeV in high multiplicity p-Pb collisions at $\sqrt{s_{NN}}=5.02$ TeV~\cite{Zhao2020}.


For  thermal partons  of $p_T>p_{T1}$  used  in  the coalescence calculations, where $p_{T1}$ is a tunable parameter in our study,   they are generated from the hadronization hyper-surface of {{\tt VISH2+1}} at $T_c=150$ MeV using the Cooper-Frye formula~\cite{Song:2010aq}. The thermal  or in-medium  masses of quarks  at this temperature  are set to $m_{u,d}=0.25$ GeV and $m_{s}=0.43$ GeV, according to the calculations from the Nambu-Jona-Lasinio model and  the Dyson-Schwinger equations~\cite{Li:2019nzj,Carroll:2008sv,Gao:2017gvf,Fu:2019hdw}. Since the thermal mass of gluons is close to  those  of quarks near $T_c$, we convert them to quark and anti-quark pairs via $gg\rightarrow q\bar{q}$, with the conversion ratio given by the abundance ratio of $u$, $d$ and $s$ quarks with thermal masses~\cite{Levai:1997yx,Silva:2014sea}. From the hadronization hyper-surface of {{\tt VISH2+1}}, we also allow soft hadrons of low $p_T$ to be  produced with the standard Cooper-Frye formula. To avoid double counting the thermal contribution to soft hadron production, we only allow thermal partons with transverse momenta above $p_{T1}$ to take part in the coalescence process, and only count those thermal mesons and baryons with transverse momenta below 2$p_{T1}$ and 3$p_{T1}$, respectively, emitted from hydrodynamics.

\begin{figure*}[ht]
\centering
\includegraphics[width=0.75\textwidth]{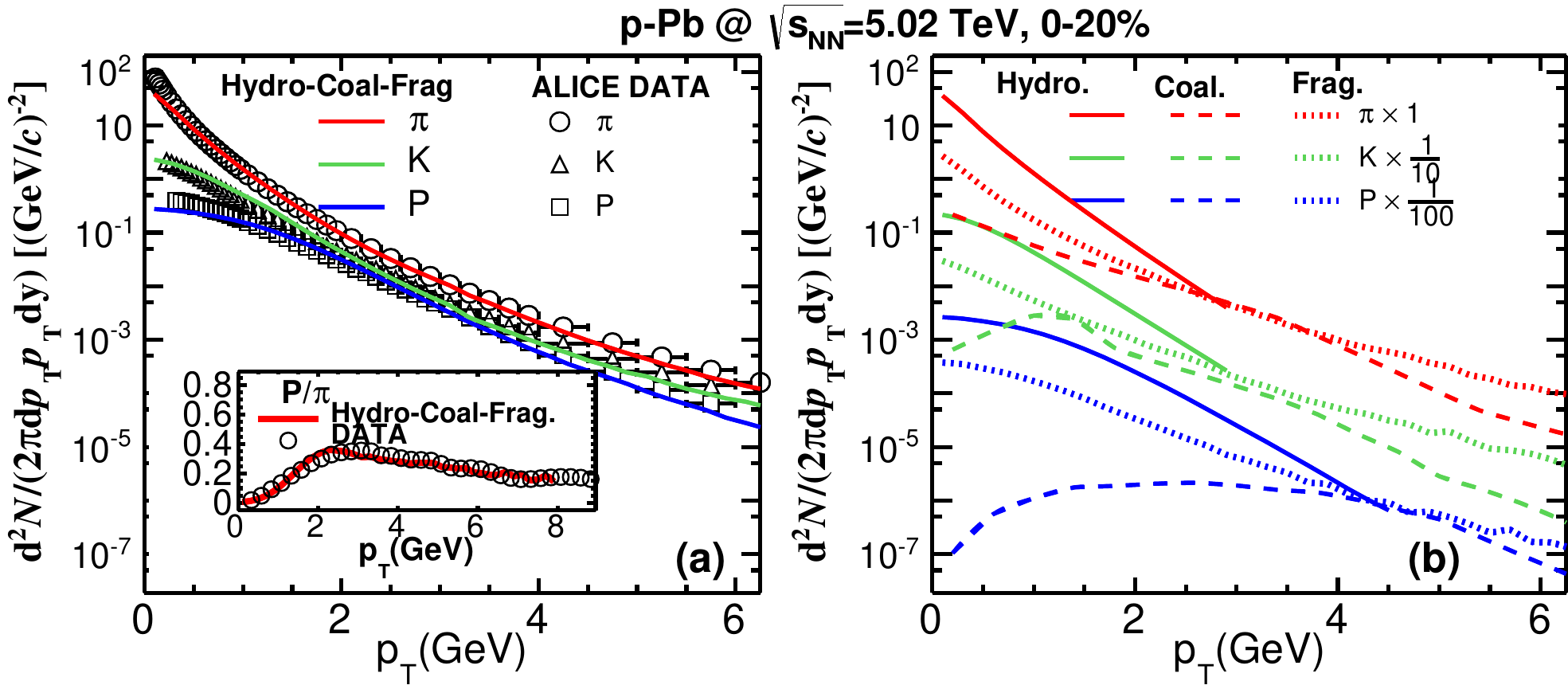}
\caption{(Color online) Left  window:  $p_T$ spectra of pions, kaons and protons in 0-20\% p-Pb   at $\sqrt{s_{NN}}=5.02$ TeV, calculated  with  the {\tt Hydro-Coal-Frag} model. The inset panel shows the proton-to-pion ratio. Right window:  contributions from hydrodynamically expanding fluid (solid lines), quark coalescence (dashed lines) and fragmentation of quenched jets (dotted lines).  The ALICE data is taken from~\cite{Adam:2016dau}.}
\label{fig:distrov2v3}
\end{figure*}

For the distributions of hard partons used in the coalescence calculations, we first generate the initial jet shower partons from PYTHIA8~\cite{Sjostrand:2007gs} and let them  move  freely until the start of hydrodynamics when they are inputted into the {{\tt LBT}} model~\cite{Wang:2013cia,He:2015pra,Cao:2016gvr,Cao:2017hhk,Chen:2017zte,Luo:2018pto} to simulate their energy losses due to elastic and inelastic interactions with the thermal partons in the evolving medium described by {{\tt VISH2+1}}. Setting the strong coupling constant $\alpha_s$ to $\alpha_s=0.15$, which is the only parameter in {{\tt LBT}},  we obtain the hard partons of $p_T>p_{T2}$ after the energy loss with $p_{T2}$ being another tunable parameter in our hybrid model. The {{\tt LBT}} model with this value of $\alpha_s$ has been shown to give a good description of the $R_{AA}$ and anisotropic flow of light and heavy flavors in the high $p_T$ region of AA collisions at RHIC and LHC~\cite{He:2015pra,Cao:2017hhk}. For the virtualities of hard gluons, which are taken to be zero in  {\tt LBT}, we follow the approach of Ref.~\cite{Han:2016uhh} by letting them to have  the values between 2$m_s$ and $m_{\rm max}$, where $m_{ \rm max}$ is the third turnable parameter in our model.   We then let these gluons to decay isotropically into $q\bar{q}$ pairs with $u\bar{u}$ and $d\bar{d}$ pairs having equal decay weight and the ratio of light to strange quarks given by the phase space and the vector nature of gluons.


After the  thermal-thermal, thermal-hard and hard-hard parton recombinations, the remnant hard partons not used in the coalescence are grouped into strings and fragmented to hard hadrons using  the ``hadron standalone mode" of PYTHIA8~\cite{Sjostrand:2007gs}\footnote{Here, we neglect the remnant thermal partons due to their very small number and negligible contribution to  hadron production at intermediate $p_T$.}.  These intermediate $p_T$ hadrons as well as  low $p_T$ thermal hadrons from hydrodynamics  and high $p_T$ hadrons from jet fragmentations are fed into {\tt UrQMD}~\cite{Bass:1998ca,Bleicher:1999xi} for the subsequent evolution of hadronic matter.

\noindent \textsl{3. Results and discussions.}
 For the values of the three parameters $p_{T1}$, $p_{T2}$, and $m_{\rm max}$ in the {\tt Hydro-Coal-Frag} model, we take them to be $p_{T1}=1.6$ GeV, $p_{T2}=2.6$ GeV and $m_{\rm max}=1.5$ GeV  the  ALICE  data from 0-20\% p-Pb collisions at $\sqrt{s_{NN}}=5.02$ TeV, where the average charged particle multiplicity is $\left<dN_{ch}/dy\right>= 35.6\pm0.8$ $(|\eta_{lab}|<0.5)$~\cite{Abelev:2013haa}. As shown in the  left  window of Fig.~1, our model  nicely reproduces the $p_T$-spectra  of  pions, kaons and protons  as well as the  peak structure in the  proton-to-pion ratio  at $p_T\approx 3$ GeV  (inset panel)  measured by ALICE.   The  right window of Fig.~1  further shows that  hadrons of momenta below 2 GeV and above 4-5 GeV are dominantly produced from hydrodynamics and jet fragmentation, respectively, while those at intermediate $p_T$ between 2-4 GeV are produced from both quark coalescence and jet fragmentation. These three $p_T$ regions  merge smoothly to provide a nice description of the $p_T$-spectra of these identified hadrons from 0 to 6 GeV.

\begin{figure*}[ht]
\centering
\includegraphics[width=1.0\textwidth]{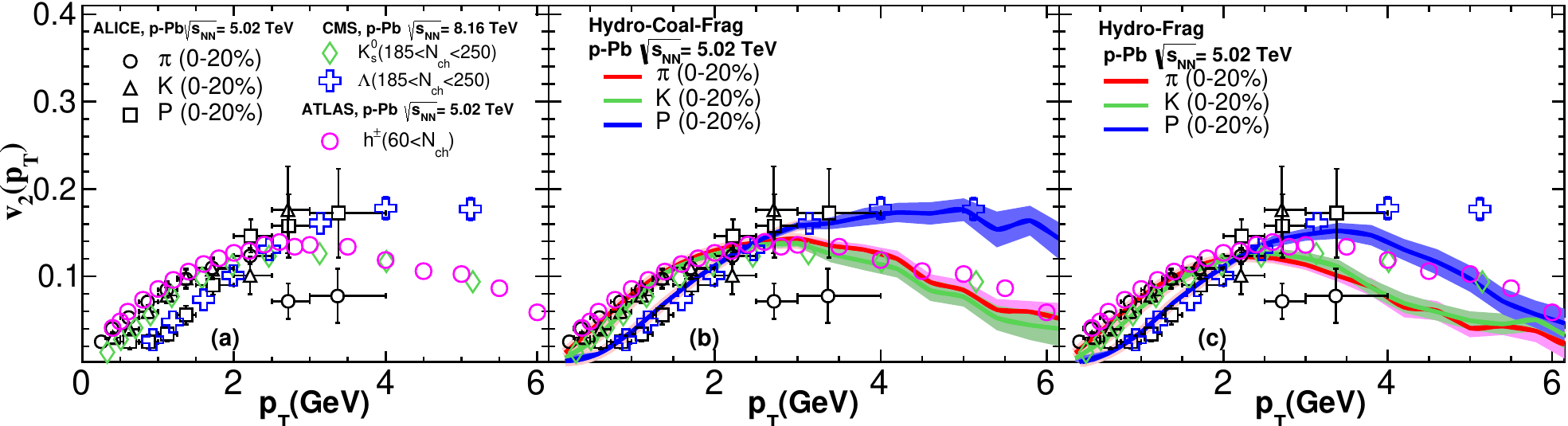}
\caption{(Color online) Differential elliptic flow $v_2(p_T)$ of  pions, kaons and protons in 0-20\%   p-Pb   at $\sqrt{s_{NN}}=5.02$ TeV measured in experiments (left window) and calculated  with  the {\tt Hydro-Coal-Frag} model (middle window) and the {\tt Hydro-Frag} model (right  window).  The ALICE, CMS and ATLAS data are taken from~\cite{ABELEV:2013wsa}, ~\cite{Sirunyan:2018toe} and~\cite{Aaboud:2016yar}, respectively. }
\label{fig:distrov2v3}
\end{figure*}
\begin{figure*}[ht]
\centering
\includegraphics[width=1.0\textwidth]{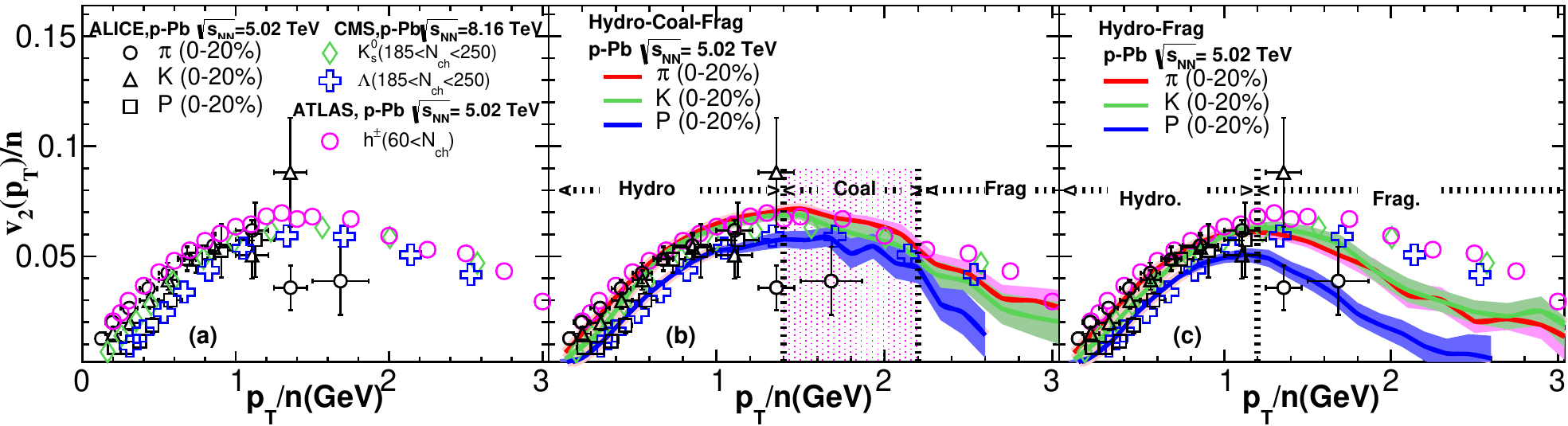}
\caption{(Color online) Similar to Fig.~2, but for the number of constituent quark scaled $v_2(p_T)$ of pions, kaons and protons. }
\label{fig:distrov2v3}
\end{figure*}

In Fig.~2, we show the $v_2(p_T)$ of pions, kaons and protons in 0-20\% p-Pb collisions at $\sqrt{s_{NN}}=5.02$ TeV from  {\tt Hydro-Coal-Frag} (middle window) and {\tt Hydro-Frag}  (right window) with  and without the coalescence contributions, respectively.  For clarity, the left window shows the experimental data~\cite{ABELEV:2013wsa,Sirunyan:2018toe,Aaboud:2016yar} without  theoretical curves.   Because of the large uncertainties of  available  data on the $v_2(p_T)$  of kaons and protons   for $p_T>2$ GeV and  the lack of $v_2(p_T)$  data above $p_T>4$ GeV  from ALICE~\cite{ABELEV:2013wsa}, we also include the ATLAS data for charged hadrons in p-Pb collisions at $\sqrt{s_{NN}}=$5.02 TeV and  the CMS data for $K_s^0$ and $\Lambda$ in p-Pb collisions at $\sqrt{s_{NN}}=$8.16 TeV.   At low $p_T$, the $v_2(p_T)$ of $K_s^0$ and $\Lambda$  from CMS  almost overlap with those of $K$ and $ p$ from ALICE.   Also, the $v_2(p_T)$ of  charged hadrons from ATLAS almost overlaps with that of pions from ALICE at low $p_T$ and with that of kaons from CMS up to $p_T=$ 5 GeV.  As pointed out in Ref.~\cite{Fries:2008hs}, the crossing of meson $v_2(p_T) $ and baryon $v_2 (p_T) $ at  $p_T\approx$  2 GeV is an indication of the transition from the hydrodynamic elliptic flow to the coalescence elliptic flow. The middle window shows that  the  {\tt Hydro-Coal-Frag} model can nicely describe the measured $v_2(p_T)$ of pions, kaons and protons from 0 to 6 GeV.  At low $p_T$, the mass ordering of $v_2(p_T)$ of identified hadrons is well reproduced by the related hydrodynamic part. For $p_T >$ 2.5 GeV,  the $v_2(p_T)$ of protons becomes larger than that of pions and kaons, which can be reproduced by the quark coalescence contributions in  {\tt Hydro-Coal-Frag}. In contrast, the {\tt Hydro-Frag} model without  the  coalescence process fails to simultaneously describe the  $p_T$ spectra and $v_2(p_T)$ of identified hadrons at $3<p_T<6$ GeV no matter how we tune the parameters.  It is worth to point out that the enhanced hadron $v_2$ at $p_T\approx 5-6$ GeV in the  {\tt Hydro-Coal-Frag} model as shown in the middle window of Fig. 2 is  due to the  order of magnitude larger  $v_2$  for  hadrons  from coalescence than  for  those from fragmentation, although more hadrons at these momenta are produced from fragmentation than from coalescence~\cite{Zhao2020}.

Figure~3 shows the  NCQ scaled elliptic flow $v_2(p_T)/n$ (with $n=2$ for mesons and $n=3$ for baryons) in 0-20\% p-Pb collisions, measured in experiments, calculated with  {\tt Hydro-Coal-Frag} and {\tt Hydro-Frag}. We again include the data from ALICE~\cite{ABELEV:2013wsa}, CMS~\cite{Sirunyan:2018toe} and ATLAS~\cite{Aaboud:2016yar} as in Fig.~2.  Although the ALICE data has large uncertainties for $ p_T/n$ larger than 1 GeV, those from CMS and ATLAS measurements show an approximate NCQ scaling of $v_2$ at intermediate $ p_T$.  With the inclusion of the quark coalescence contribution, results from the {\tt Hydro-Coal-Frag} model show an approximate scaling behavior between 1.4$<p_T/n<$2.2 GeV, as observed in experiments, even though contributions from resonance decays and jet fragmentation can lead to a slightly violation of the NCQ scaling of $v_2$.  In contrast, the results from the {\tt Hydro-Frag} model without the quark coalescence process not only underestimates the magnitude of $v_2(p_T)/n$ but also violates the NCQ scaling behavior at intermediate $p_T$.

We note that using a smaller strong coupling constant than the value $\alpha_s=0.15$ adopted in the early study~\cite{He:2015pra,Cao:2017hhk} would increase the $p_T$ spectra and decrease the differential elliptic flow $v_2(p_T)$ of hadrons with $p_T>3$ GeV, resulting thus in a poor description of the experimental data.  These effects become, however, negligible for $p_T>8$ GeV, consistent with the expectations that no obvious signatures due to the energy loss of energetic jets have been  seen in small collision systems~\cite{ATLAS:2014cza,ALICE:2012mj,Khachatryan:2015xaa,Acharya:2018qsh,Acharya:2017okq,Albacete:2013ei,Eskola:2016oht,Abelev:2014hha,Khachatryan:2015uja,Dong:2019byy,Khachatryan:2015sva,Adam:2016ich,Du:2018wsj}. Also, using  a smooth transition in the $p_T$ spectra for thermal and hard partons gives essentially the same results  shown in the present study from a sharp cut in  both {\tt Hydro-Coal-Frag} and {\tt Hydro-Frag} models. Details on these results and the sensitivities of the results presented in the present study to the values of the parameters $p_{T1}$, $p_{T2}$ and $m_{\rm max}$ in the {\tt Hydro-Coal-Frag} model will be reported elsewhere~\cite{Zhao2020}.

\noindent \textsl{4. Summary.} In this paper, we have carried out the first quantitative and timely study of the NCQ scaling of elliptic flow $v_2$ of pions, kaons and protons at intermediate $p_T$ in high multiplicity p-Pb collisions at $\sqrt{s_{NN}}=$ 5.02 TeV via the coalescence of soft thermal partons from the {\tt VISH2+1} hydrodynamics and hard partons from the energy loss  {\tt LBT} model.   Adding low $p_T$ hadrons from the hydrodynamically expanding fluid and high $p_T$ hadrons from jet fragmentation  to these intermediate $p_T$ hadrons from quark coalescence,  our  {\tt Hydro-Coal-Frag} hybrid model can simultaneously describe the $p_T$-spectra and differential elliptic flow $v_2(p_T)$ of identified hadrons over the $p_T$ range from 0 to 6 GeV.  We have also demonstrated that the inclusion of the quark coalescence contribution to the production of hadrons  is essential in reproducing the measured $v_2(p_T)$ of these identified hadrons and their observed approximate NCQ scaling at intermediate $p_T$. Results from the present study thus provides a strong indication for the existence of the partonic degrees of freedom and the possible  formation of the QGP in high multiplicity p-Pb collisions at $\sqrt{s_{NN}}=5.02$ TeV. \\

\noindent \textsl{Acknowledgments.} We thank P. Bo$\rm\dot{z}$ek, S. Cao, P. Christiansen, H. Elfner,  Y. He, J.-Y. Jia, T. Luo, B. Schenke,  T. Sj$\rm\ddot{o}$strand, X.-N. Wang, W.-J. Xing, and J.-X. Zhao  for helpful discussions. W. Z., Y. L. and H. S. are supported by the NSFC and the MOST under grant Nos.  11435001 and 11675004. G.-Y. Q. is supported by NSFC under Grants No. 11775095, 11890711 and 11375072, and by CSC under Grant No. 201906775042.  C.M. K. is supported by US DOE under Contract No. DE-SC0015266 and the Welch Foundation under Grant No. A-1358. W. Z. and H. S. also acknowledge the computing resources provided by the SCCAS and Tianhe-1A Super-computing Platform in China and the High-performance Computing Platform of Peking University.

\end{document}